\documentclass[pdflatex,sn-nature]{sn-jnl}

\usepackage{booktabs}
\usepackage{lineno}
\usepackage{lmodern}
\usepackage{hyperref}
\usepackage{bookmark}

\usepackage{amsmath}
\usepackage{amssymb}
\usepackage{enumitem}
\usepackage{url}
\usepackage{bm}

\setenumerate[1]{itemsep=0pt,partopsep=0pt,parsep=\parskip,topsep=5pt}
\setitemize[1]{itemsep=0pt,partopsep=0pt,parsep=\parskip,topsep=5pt}
\setdescription{itemsep=0pt,partopsep=0pt,parsep=\parskip,topsep=5pt}

\newcommand{\vect}[1]{{\bm{#1}}}				
\newcommand{\matx}[1]{{\uppercase{\bm{#1}}}}	
\newcommand{\opt}[1]{{\mathcal{#1}}}		
\newcommand{\vn}{\vect{n}}	
\newcommand{\vx}{\vect{x}}	
\newcommand{\vy}{\vect{y}}   
\newcommand{\vtheta}{\vect{\theta}}   
\newcommand{\mB}{\matx{B}}   
\newcommand{\mI}{\matx{I}}   
\newcommand{\oF}{\opt{F}}   

\newcommand{\remove}[1]{ }

\newcommand{\arcsec}{^{\prime \prime}}

\begin{document}



\title[Astronomical Image Denoising]{Astronomical Image Denoising by Self-Supervised Deep Learning and Restoration Processes}


\author*[1,2,5,6]{\fnm{Tie} \sur{Liu}}\email{liu\_tie\_zimu@nju.edu.cn}

\author*[3,4]{\fnm{Yuhui} \sur{Quan}}\email{csyhquan@scut.edu.cn}

\author*[5,6]{\fnm{Yingna} \sur{Su}}\email{ynsu@pmo.ac.cn}

\author*[1,2]{\fnm{Yang} \sur{Guo}}\email{guoyang@nju.edu.cn}

\author[7]{\fnm{Shu} \sur{Liu}}\email{yolandaliu54@163.com}

\author[5,6]{\fnm{Haisheng} \sur{Ji}}\email{jihs@pmo.ac.cn}

\author[1,2]{\fnm{Qi} \sur{Hao}}\email{haoqi@nju.edu.cn}

\author[1,2]{\fnm{Yulong} \sur{Gao}}\email{yulong@nju.edu.cn}

\author[8,9]{\fnm{Yuxia} \sur{Liu}}\email{1901111599@pku.edu.cn}

\author[1,2]{\fnm{Yikang} \sur{Wang}}\email{wyk@nju.edu.cn}

\author[10]{\fnm{Wenqing} \sur{Sun}}\email{sunwenqing@nbu.edu.cn}

\author[1,2]{\fnm{Mingde} \sur{Ding}}\email{dmd@nju.edu.cn}

\affil*[1]{\orgdiv{School of Astronomy and Space Science}, \orgname{Nanjing University}, \orgaddress{\street{163 Xianlin Avenue}, \city{Nanjing} \postcode{210023}, \state{Jiangsu}, \country{P.R. China}}}

\affil*[2]{\orgdiv{Key Laboratory of Modern Astronomy and Astrophysics(Nanjing University)}, \orgname{Ministry of Education}, \orgaddress{\street{163 Xianlin Avenue}, \city{Nanjing} \postcode{210023}, \state{Jiangsu}, \country{P.R. China}}}

\affil*[3]{\orgdiv{School of Computer Science and Engineering}, \orgname{South China University of Technology}, \orgaddress{\street{382 Waihuan East Road}, \city{Guangzhou} \postcode{510006}, \state{Guangdong}, \country{P.R. China}}}

\affil[4]{\orgname{Pazhou Lab}, \orgaddress{\street{248 Qiaotou Street}, \city{Guangzhou} \postcode{510335}, \state{Guangdong}, \country{P.R. China}}}

\affil[5]{\orgdiv{Key Laboratory of Dark Matter and Space Astronomy, Purple Mountain Observatory}, \orgname{Chinese Academy of Sciences}, \orgaddress{\street{10 Yuanhua Road}, \city{Nanjing} \postcode{210034}, \state{Jiangsu}, \country{P.R. China}}}

\affil[6]{\orgdiv{School of Astronomy and Space Science}, \orgname{University of Science and Technology of China}, \orgaddress{\street{96 Jinsai Road}, \city{Hefei} \postcode{230026}, \state{Anhui}, \country{P.R. China}}}

\affil [7]{\orgdiv{Department of Rheumatology}, \orgname{Nanjing Drum Tower Hospital, The Affiliated Hospital Of Nanjing University}, \orgaddress{\street{321 Zhongshan Road}, \city{Nanjing} \postcode{210008}, \state{Jiangsu}, \country{P.R. China}}}

\affil [8]{\orgdiv{College of Engineering}, \orgname{Peking University}, \orgaddress{\street{5 Yiheyuan Road}, \city{Beijing} \postcode{100871}, \country{P.R. China}}}

\affil [9]{\orgdiv{State Key Laboratory of Low-carbon Smart Coal-fired Power Generation and Ultra-clean Emission, China Energy Science and Technology Research Institute Co., Ltd}, \orgaddress{\street{10 Xianjing Road}, \city{Nanjing} \postcode{210034}, \country{P.R. China}}}

\affil [10]{\orgdiv{Faculty of Electrical Engineering and Computer Science}, \orgname{Ningbo University}, \orgaddress{\street{No. 818, Fenghua Road, Ningbo},  \postcode{315211}, \city{Zhejiang}, \country{P.R. China}}}

\abstract{Image denoising based on deep learning has witnessed significant advancements in recent years. However, existing deep learning methods lack quantitative control of the deviation or error on denoised images. The neural networks Self2Self is designed for denoising single-image, training on it and denoising itself, during which training is costly. In this work we explore training Self2Self on an astronomical image and denoising other images of the same kind, which is suitable for quickly denoising massive images in astronomy. To address the deviation issue, the abnormal pixels whose deviation exceeds a predefined threshold are restored to their initial values. The noise reduction includes training, denoising, restoring and named TDR-method, by which the noise level of the solar magnetograms is improved from about 8 G to 2 G. Furthermore, the TDR-method is applied to galaxy images from the Hubble Space Telescope and makes weak galaxy structures become much clearer. This capability of enhancing weak signals makes the TDR-method applicable in various disciplines.}



\maketitle
\section{Introduction}\label{sec1}

Astronomy has entered the big data era, with the advancements of ground-based and space-borne telescopes. Massive digital images are being collected and denoising these images becomes increasingly important. Taking solar physics as an example, the Solar Dynamics Observatory (SDO\cite{Pesnell2012}) is a space-borne telescope that plays a significant role in collecting data, with its higher-level derivatives exceeding 7 PB\cite{Schrijver2016IAUTA}. Additionally, ground-based telescopes such as the Goode Solar Telescope (GST\cite{Cao2010}) and the New Vacuum Solar Telescope (NVST\cite{Liu2014}) also contribute significantly to the wealth of solar images. In the field of galaxies and cosmology, the Hubble Space Telescope (HST) is renowned for providing high-quality images. Since its launch in 1990, the HST has conducted more than 1.5 million observations.

The aforementioned astronomical instruments can be simplified into two parts: light focusing systems and charge coupled device (CCD) cameras. All astronomical images obtained by these instruments inevitably contain noise, mainly from imperfections in the light focusing systems and CCD cameras. Additive White Gaussian Noise (AWGN), impulse noise (Salt-and-pepper noise), quantisation noise and Poisson noise are frequently studied\cite{Buades2005}. AWGN primarily originates from analog circuitry; quantisation noise, impulse noise, and Poisson noise, mainly result from manufacturing defects, bit errors, and insufficient photon counts during image acquisition and transmission\cite{GOYAL2020220}.

Researchers have made continuous efforts to develop denoising methods. The most common approaches are spatial domain and transform domain methods. Spatial domain methods exploit the similarities between pixels or patches to perform denoising. The Gaussian filter, for example, is widely used to reduce high-frequency signals\cite{Seddik2012IJCA}. Transform domain methods involve transforming images into the frequency domain and analyzing the similarities and dissimilarities of transformed coefficients. The Fourier transform is a widely used technique. Moreover, based on different image features and task requirements, other denoising techniques such as anisotropic diffusion methods, hybrid
methods and morphological analysis are also frequently used\cite{Elad20063736,YANG2014152,GOYAL2020220}.

Recently, deep learning based methods have emerged and shown remarkable denoising effects that surpass classic methods. These methods can be categorized into two types: supervised learning\citep[e.g.,][]{burger2012image,vemulapalli2016deep,zhang2017beyond,Park_2020} and unsupervised learning based methods\citep[e.g.,][]{lehtinen2018noise2noise,azBaso2019}. The former learns the denoiser on clean-noisy image pairs, the latter does not require clean images. Representative noise reduction works based on supervised learning includes: U-net\cite{Ronneberger2015} for image enhancement and denoising\cite{Chen2018LearningTS,Vojtekova2020,Qi2022AnIU}, generative adversarial networks (GAN)\cite{goodfellow2014generative} to recover features of degraded images and denoising\cite{Schawinski2017MNR,Chen2018ImageBD,Park_2020}. One key factor that contributes to the performance of these supervised deep learning based methods is the availability of massive clean-noisy image pairs for training. However, collecting a substantial amount of such pairs can be challenging and expensive.

With further researches, unsupervised learning based methods has emerged. The neural network Noise2Noise first show the possibility of restoring images by training on corrupted images alone\cite{lehtinen2018noise2noise}. Similar works such as Noise2Void\cite{Krull2019} and Noise2Self\cite{batson2019noise2self} have further explored learning denoisers on noisy images. The application\cite{azBaso2019} of Noise2Noise to images from the Swedish 1-m Solar Telescope, has revealed the potential to recover weak signals. To achieve good performance, the aforementioned unsupervised learning based methods typically require training on a sufficient number of external images that are highly related to the image being processed. However, collecting such images is challenging. To address this challenge, the neural network Self2Self\citep{Quan_2020_CVPR} is designed. Unlike other unsupervised methods, Self2Self can be trained on a single image and achieves even better denoising results, which makes Self2Self suitable for scenarios where image collection is limited. Due to the fact that the training data only contains noisy images themselves, denoising by Self2Self is a self-supervised learning based method, which belongs to unsupervised deep learning. However, unsupervised learning based methods are susceptible to produce abnormal results due to the lack of supervision.

The noise reduction of astronomical images not only aims to make the images cleaner, but also emphasizes preserving the accuracy of each pixel. To ensure image accuracy, the change of each pixel in a denoised image should not exceed the initial noise level. Based on Self2Self, we propose a general denoising method called TDR-method which consists of three operations: 1. training (T) on a single-image, 2. denoising (D) other images of the same kind, 3. restoring (R) abnormal pixels. The TDR-method allows for controlling the accuracy of the denoised images and is suitable for denoising scientific data of various disciplines and fields. Astronomical images are used to verify the denoising effect of this method.

\section{Results} \label{sec: Results0}

\subsection{Denoising and evaluation on HMI magnetograms}\label{sec: Results}

\begin{figure*}
	\centering \includegraphics[width=\textwidth]{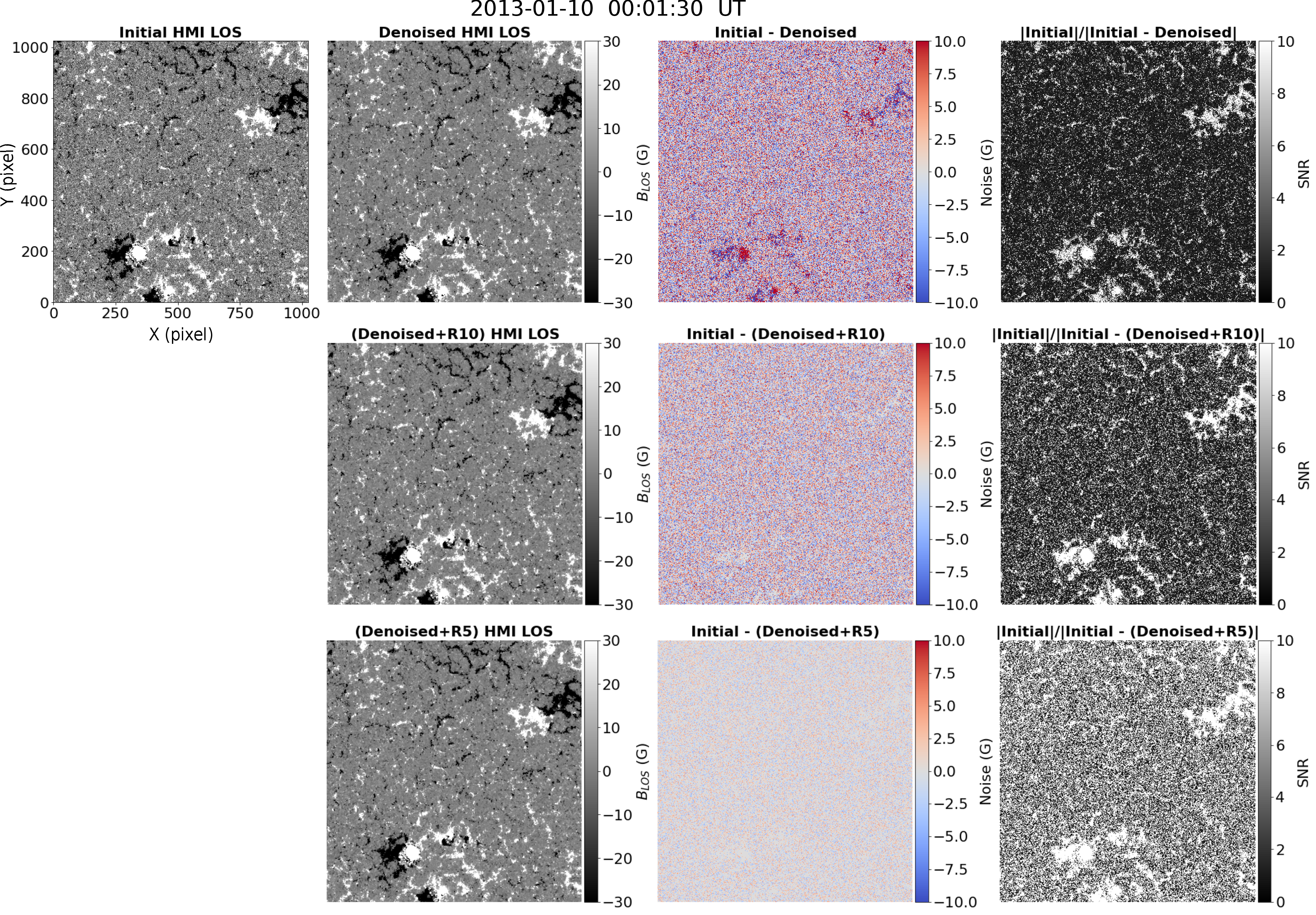}
	\caption{Comparisons between the initial and denoised HMI LOS magnetograms. We show the initial magnetogram observed at 00:01:30 UT on 2013 January 10 in the first column. The primary denoised image by Self2Self is titled `Denoised HMI LOS'. The senior denoised images with restoring operations are titled `(Denoised+R10) HMI LOS' for $R=10$ G and `(Denoised+R5) HMI LOS' for $R=5$ G. The third column shows the residual images (the difference maps obtained by subtracting the denoised image from the initial one) and the signal-to-noise ratio (SNR) images are displayed in the last column. Note that the pixels where Initial=Denoised are extremely rare, and excluded on the SNR images.}
	\label{fig: f1} 
\end{figure*}

\begin{figure*}
	\centering \includegraphics[width=\textwidth]{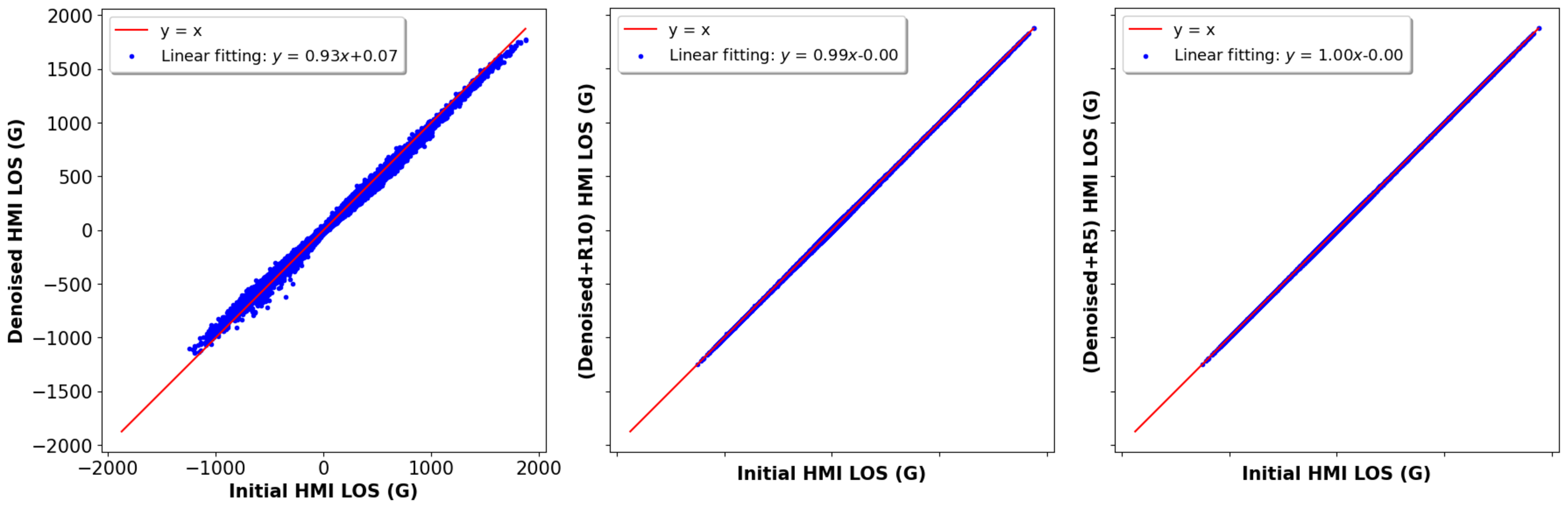}
	\caption{The scatter plots of the initial and denoised magnetograms in Figure \ref{fig: f1}. The pixel values of the initial and denoised magnetograms are scattered along horizontal and vertical axes, respectively. A line ($y=x$ ) is displayed for reference. We fit the dots with a linear function ($y=a \ x+b$) and the parameters $a$ and $b$ are listed on the top-left corner. } 
	\label{fig: f2} 
\end{figure*}

\begin{figure*}
	\centering \includegraphics[width=\textwidth]{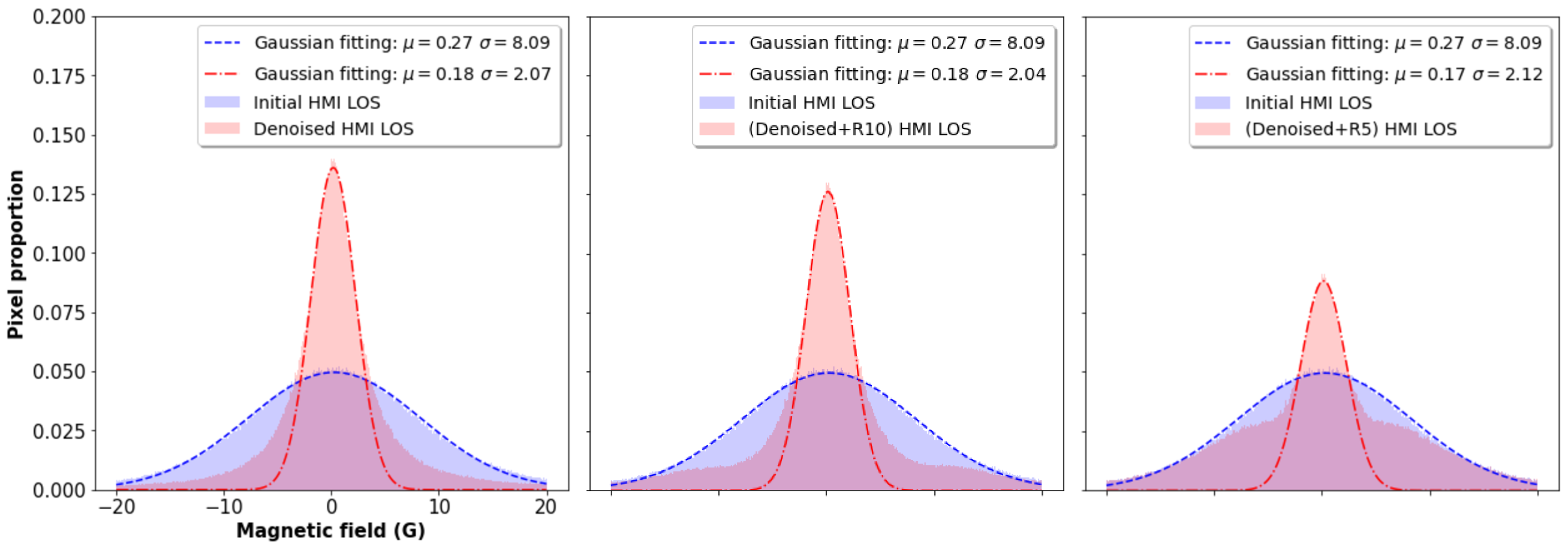}
	\caption{Gaussian fittings of the initial and denoised magnetograms in Figure \ref{fig: f1}. The blue and red regions show the distributions of the pixel values in the range of $[-20 G, 20 G]$. The corresponding Gaussian fittings are marked by the blue dashed curve and red dash-dotted curve, respectively. The mean values $\mu$ and standard deviations $\sigma$ are listed in the top-right corner.} 	
	\label{fig:f3}  
\end{figure*}

\begin{figure*}
	\centering \includegraphics[width=\textwidth]{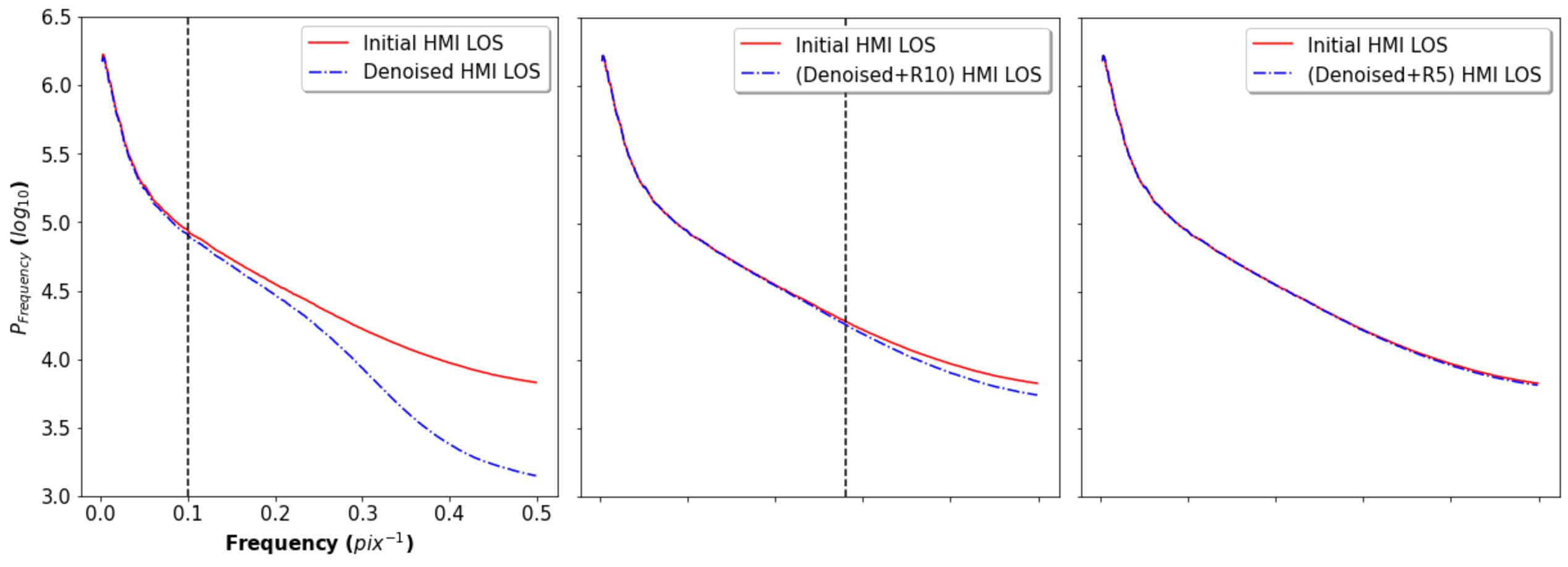}
	\caption{The spatial Fourier power spectrum of the initial and denoised magnetograms in Figure \ref{fig: f1}. The vertical dashed line marks the location $x$ where the two curves start to deviate from each other, $x=0.10$ for the first panel and $x=0.28$ for the second panel.} 	
	\label{fig:f4}  
\end{figure*}

We collect ten line-of-sight (LOS) magnetograms from the Helioseismic Magnetic Imager (HMI\cite{Schou2012solphys229259}) to train and test the neural networks Self2Self. First, the ten magnetograms are divided into the training group (containing one magnetogram) and denoising group (containing the other nine magnetograms). Second, considering that each of the 10 magnetograms can serve as the training image, we conduct 10 different training and denoising experiments in which each of the 10 magnetograms is used as the training image and the other nine as the denoising images. We find that the denoising effects of the ten experiments are almost the same, which proves that denoising by Self2Self is robust.

In Figure \ref{fig: f1}, we show comparisons of the initial and denoised HMI LOS magnetograms. The denoised images come from the first training strategy with the first image as the training sample. From the first row, we find that the primary denoised magnetogram (titled `Denoised HMI LOS') becomes much cleaner and is highly consistent with the initial magnetogram. The residual images represent the absolute error and show that the noise is randomly distributed, and the amplitude of the noise is correlated with the signal strength. However, we find that in the regions with strong magnetic field, the positive field is underestimated (red) and the negative field is overestimated (blue). Considering the absolute values, the strong positive and negative magnetic fields are both underestimated. The average value of the underestimation is about 10 G, which is in the same scale as the noise level\citep{Liu2012SoPh}. The underestimations can have an impact on quantitative analyses, such as calculations of magnetic flux and oscillations\cite{ji2021RAA}. The last image in the first row of Figure \ref{fig: f1} shows that the signal-to-noise ratio (SNR) is high in regions with strong magnetic field, whereas the SNR in other regions is relatively unsatisfactory, due to that the weak magnetic field is almost equivalent to the noise. 

It is difficult to solve the underestimation by adjusting the neural networks and training strategies, because unsupervised learning based methods are difficult to control. In a different train of thought, we propose a simpler solution. First finding the abnormal pixels whose changes of absolute values greater than a threshold $R$ after denoising, second restoring the abnormal pixels to their initial values. By this restoring operation, the senior denoised images are obtained and titled `(Denoised+R10) HMI LOS' for $R=10$ G and `(Denoised+R5) HMI LOS' for $R=5$ G in the second and third rows of Figure \ref{fig: f1}. The deviation between the value of every pixel in the magnetogram before and after the noise reduction will not exceed $10$ G for `(Denoised+R10) HMI LOS' and $5$ G for `(Denoised+R5) HMI LOS'. The corresponding residual images show that the underestimation is effectively controlled. The proportions of the restored pixels are $13.98\%$ for `(Denoised+R10) HMI LOS' magnetograms and $44.45\%$ for `(Denoised+R5) HMI LOS' magnetograms. To ensure the accuracy of the denoised images, the value of R should be smaller than the noise level. So $R=5$ G is acceptable and suitable for the HMI LOS magnetograms.

We plot the pixel values of the initial and denoised HMI LOS magnetograms in Figure \ref{fig: f1} along horizontal and vertical axes in Figure \ref{fig: f2} and fit the scattered points with linear functions. The points in Figure \ref{fig: f2} show high correlation, with linear coefficients (LC) of 0.93, 0.99 and 1.00, respectively. We note that the scatter plot also shows the underestimation which is indicated by the slight deviation of the two ends from the red line in the first panel. While the second and third panels show that the underestimation is well reduced in the `(Denoised+R10) HMI LOS' and `(Denoised+R5) HMI LOS' magnetograms. The absolute error (AE), mean square error (MSE), pixel-to-pixel Pearson Correlation Coefficient (PCC), LC, average flux of residual images (AFR) and average absolute flux of residual images (AAFR) are calculated. The calculation methods are described in Supplementary Information.  We list the mean values (MV) and standard deviations (SD) of AE, MSE, PCC, LC, AFR and AAFR for the nine denoised magnetograms in Supplementary Information. These quantitative evaluations with their small standard deviations suggest that the TDR-method is robust and reliable.


Figure \ref{fig:f3} shows the histograms of pixel values in the range of $\mathrm{[-20 \ G, 20 \ G]}$ on the magnetograms of Figure \ref{fig: f1}, with Gaussian fitting curves. The standard deviation of the Gaussian fitting can represent the noise level of the magnetograms\cite{Liu2012SoPh}. The noise levels are reduced from 8.09 G to 2.07 G for the `Denoised' magnetogram, to 2.04 G for the `(Denoised+R10)' magnetogram and to 2.12 G for the `(Denoised+R5)' magnetogram, which are better than the average noise level of 3.21 G achieved by a supervised learning method\cite{Park2020ApJ}. The mean value and standard deviation of the noise level for the nine magnetograms are listed in Supplementary Information. We find that the restoring operation do not excessively increase the noise level on the premise of maintaining the accuracy of the pixel value. Meanwhile, we notice that both sides of the red histogram become convex with R=10 G or R=5 G. The convex parts are caused by the restored pixels which are located inside the Gaussian fitting of the initial magnetogram and outside the Gaussian fitting of the denoised magnetogram. This indicates that the restoring operation makes a portion of weak magnetic signals between the initial noise level ($\sim 8$ G) and the new noise level ($\sim 2$ G) become distinguishable.

In Figure \ref{fig:f4}, we find that the difference between the initial and denoised magnetograms starts at the frequency of about $0.10\ pix^{-1}$ for the `Denoised' magnetogram, $0.28\ pix^{-1}$ for the `(Denoised+R10)' magnetogram and $0.50\ pix^{-1}$ for the `(Denoised+R5)' magnetogram in the power spectrum. It indicates that the denoising mainly works in scales smaller than $10 \times 10$ pixels, while in scales larger than $10 \times 10$ pixels, the denoised magnetograms are highly consistent with the initial magnetograms. However, magnetic structures smaller than $10 \times 10$ pixels may exist. By incorporating the restoring operation, we can preserve structures at significantly smaller scales, approximately $4 \times 4$ pixels for the (Denoised+R10)' magnetograms and $2 \times 2$ pixels for the (Denoised+R5)’ magnetograms.

\subsection{Scientific application to Hubble images} \label{sec: application}

\begin{figure*}
	\centering \includegraphics[width=\textwidth]{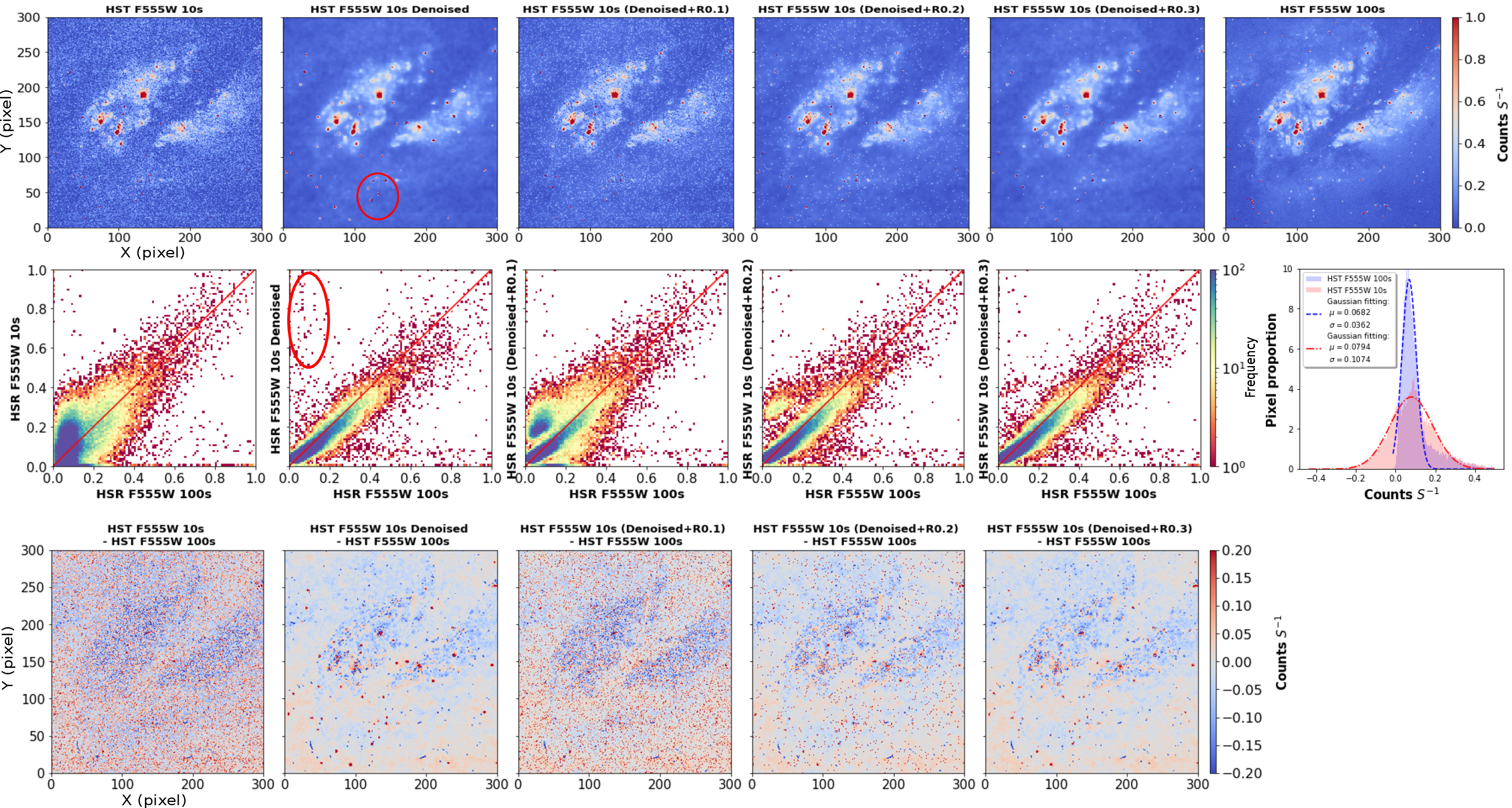}
	\caption{An example of the denoised images for the galaxy center of NGC 1365 observed by HST with the filter of F555W. The first and last panels in the upper row show images with the exposure time of 10 s and 100 s. The middle four images are the corresponding denoised images with R = $\infty$ (equivalent to the restoring operation not working), 0.1, 0.2 and 0.3, respectively. The density-scatter plots of the first five images with the last image in the upper row are displayed in the first five panels of the second row. The histograms and Gaussian fittings of the HST images with the exposure time of 10 s and 100 s are displayed in the last panel of the second row. The residual maps are displayed in the bottom row. The unit of pixel values is counts per second.} 	
	\label{fig:f5}  
\end{figure*}

\begin{figure*}
	\centering \includegraphics[width=\textwidth]{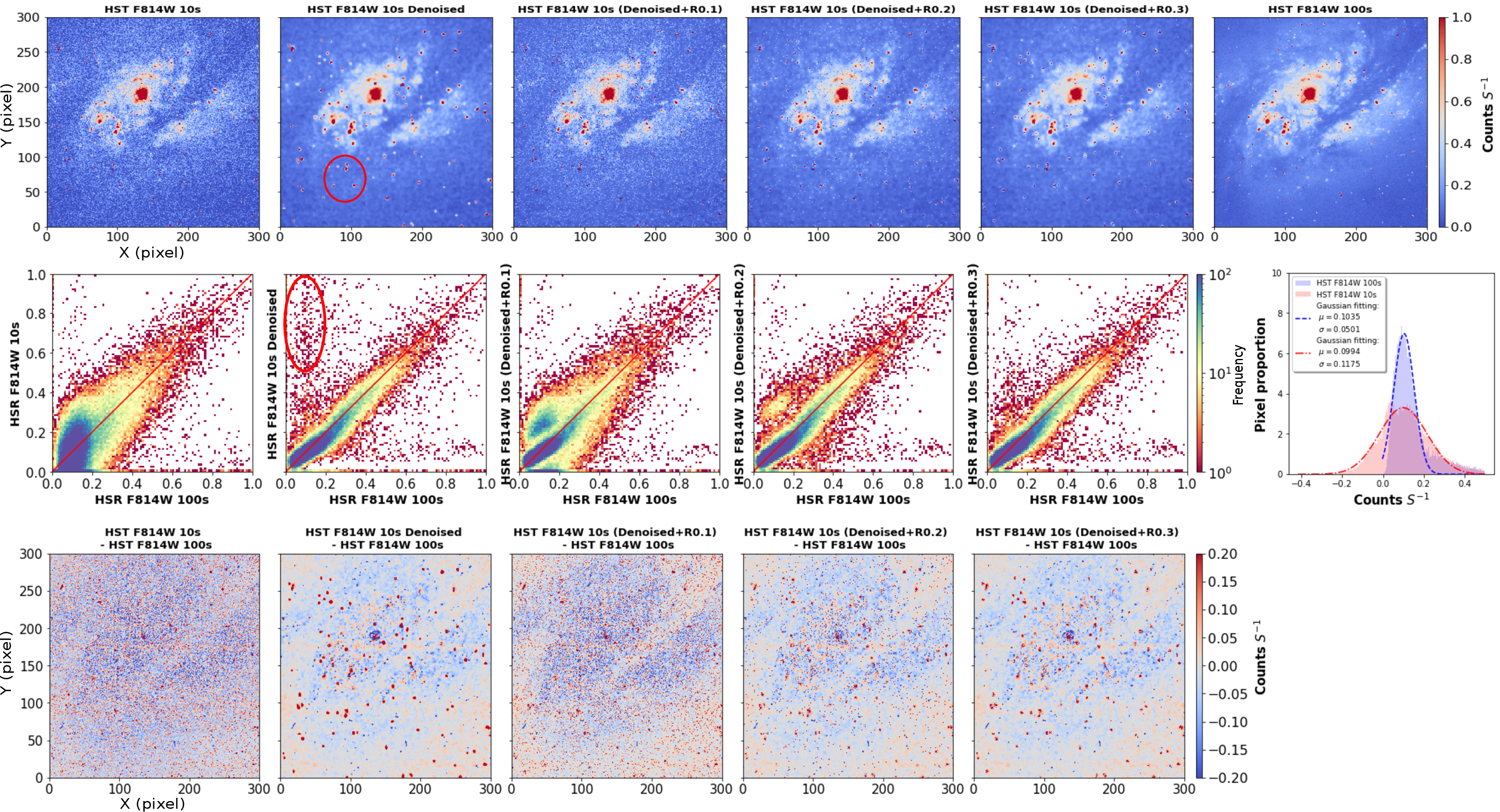}
	\caption{An example of the denoised images for the galaxy center of NGC 1365 observed by HST with the filter of F814W. The layout and meanings of every panel are the same as those in Figure \ref{fig:f5}, but for the filter of F814W.} 	
	\label{fig:f6}  
\end{figure*}

In astronomy, it is difficult to detect the faint structures of extragalactic galaxies within limited observation time even using the space telescopes, i.e. Hubble Space Telescope. Our TDR-method can also denoise the extragalactic images with limited exposure time and help to identify faint objects and structures. Here, we show an application of the TDR-method in the observation of the nearby ($z$ = 0.0054, at a distance of 21.2 Mpc, where 1$\arcsec \sim$  90 pc) galaxy NGC 1365\citep[e.g., ][]{Lindblad1999,Galliano2005,Wang2009,Gao2021}. In Figures \ref{fig:f5} and \ref{fig:f6}, we show the central region ($30 \arcsec \times 30 \arcsec$) images of NGC 1365 observed by HST with filters of F555W and F814W. As shown in the first panel of Figure \ref{fig:f5}, because of the significant noise, the galactic structures are indistinctive in the images with the exposure time of 10-second. We use the TDR-method with R = $\infty$, 0.1, 0.2 and 0.3 (the unit is counts per second) to denoise the 10-second images and show the denoised images in the middle panels. We note that the primary denoised image (R = $\infty$) become cleaner, but some abnormal pixels marked by the red circle are generated. These abnormal pixels are more obvious and marked by the red ellipse in the density-scatter plots. Figure \ref{fig:f6} indicates that the abnormal pixels occur more frequently in the filter of F814W. From the residual maps, we observe that some pixels initially exhibit abnormalities in the 10-second images, as compared to the corresponding 100-second images. Furthermore, these abnormal pixels are accentuated during the denoising process.

To remove the enhanced abnormal pixels, we adopt the restoring operation and show the senior denoised images with R = 0.1, 0.2 and 0.3 in the third, forth and fifth panels of Figures \ref{fig:f5} and \ref{fig:f6}. By comparing the corresponding density-scatter plots, we find that when R = 0.3, the best denoised images are obtained. Gaussian fittings in the last panels of the second row in Figures \ref{fig:f5} and \ref{fig:f6} indicate that the noise level of the HST images is about 0.4 or 0.5 counts per second. To meet a reasonable accuracy requirement, the R needs to be less than the noise level. So R = 0.3 is acceptable and appropriate for the HST images. 

Comparing with the 100-second HST images in the last panel (upper row) of Figures \ref{fig:f5} and \ref{fig:f6}, we find that the senior denoised images with R = 0.3 are the best and the identifications of fainter star clusters as well as dust lanes are reliable. The residual maps in Figures \ref{fig:f5} and \ref{fig:f6} show that the restoring operation can suppress abnormal pixels generated during the denoising process, but the initially existing abnormal pixels are retained as signals. Correcting these initially existing abnormal pixels relies on comparing with long exposure images, which is beyond the scope of noise reduction and may be a further step of future improvement for the TDR-method.

\section{Summary and discussion} \label{sec: Summary}

In this study, we investigate the denoising of astronomical images based on the unsupervised neural networks Self2Self. The denoising process consists of three operations: 1. training (T) Self2Self on a single-image, 2. denoising (D) other images of the same kind, 3. restoring (R) abnormal pixels according to accuracy requirements. We first test this TDR-method on HMI LOS magnetograms and find that the noise level is improved from about 8 G to 2 G. Detailed comparisons between the initial and denoised magnetograms are carried out. The result shows that the denoised magnetograms become cleaner and smoother without sacrificing temporal and spatial resolutions. During the denoising process, we notice the underestimation of the strong magnetic field in the primary denoised magnetograms and the underestimation exceeds the noise level. To address this issue, we propose a two-step solution: first, identifying abnormal pixels whose change values exceed a threshold $R$, and second restoring these pixels to their initial values. Generally, the value of $R$ should be lower than the noise level of the image. When $R=0$, the denoised images are restored to the initial images; when $R=\infty$, the restoring operation does not work. By controlling the value of $R$, we can obtain senior denoised images that meet a reasonable accuracy requirement.

The denoising effects of the TDR-method is supported by quantitative evaluations such as AE, MSE, PCC, LC, and AFR, which illustrate the robustness and reliability of the method. The denoising of HST images demonstrates the effectiveness of the TDR-method in improving the clarity and detail of galaxy structures. The denoised images with a 10-second exposure time exhibit enhanced consistency with the corresponding images of 100-second exposure time. This implies that the TDR-method can significantly reduce the noise level of 10-second HST images. The restoration process can complement the deep learning results, which inspires that just good scores of the AE, MSE, PCC, etc. are not enough. The pixel level accuracy is more important for many scientific fields. Restoring the abnormal pixels to their initial values is the simplest restoration process, we also recommend to use other different restoration processes, such as restoring to the mean values of the ambient pixels and other reliable values.

Regarding the accuracy of the denoising results, different options are adopted.

\begin{enumerate} 
\item In the situation of observational images, the clean target images are not available. If images with higher exposure times are also not available, the denoising accuracy is estimated by comparing the denoised images with the original images. The goal is to control the changes in pixel values before and after denoising to be less than the noise level and make the denoised images meet a reasonable accuracy requirement, as illustrated in Section \ref{sec: Results}.
\item If images with higher exposure times are available, the denoising accuracy is estimated by comparing the denoised images with the images of higher exposure times, as discussed in Section \ref{sec: application}.
\item If the clean target images are available, we define the denoising accuracy by comparing the denoised images with the clean target images. Please refer to Supplementary Information, which shows the denoising examples of synthetic images from RAMENS code\citep{Iijima2017ApJ} with known noises. 
\end{enumerate}

The latter two scenarios are more suitable for supervised learning based methods. In astronomy, the first scenario is common, and the TDR-method can play an irreplaceable role.

For astronomical observations, both high temporal resolution and high SNR are necessary and important but they often restrict each other. Considering different goals, to achieve higher SNR, longer integration times has to be invested; to achieve higher temporal resolutions, higher SNR has to be sacrificed. Taking the HST as an example, ultra-deep images are usually obtained by using long exposure times, for example, several days\cite{Beckwith2006AJ}. Taking advantage of the TDR-method, we propose a two-step strategy to achieve both high temporal resolutions and high SNR. First observing the objects with a higher temporal resolution to get images with lower SNR, and second improving the SNR by the TDR-method. This is especially beneficial for astronomical sky survey projects\citep[e.g., ][]{Su2019ApJS,Norris2021PASA} that require abundant observation times. The TDR-method holds promising potential for denoising and improving imaging efficiency in diverse disciplines. 
\section{Methods}\label{secA}
\subsection{Basic ideas}
Image denoising aims at recovering the clear version of a corrupted image, which is widely used in many applications. In general, image denoising considers estimating the clear image $\vx$ from the corrupted image $\vy$ generated by the following process: $\vy=\vx+\vn$, where $\vn$ denotes the noise distribution.

Inspired by the effectiveness of blind-spot methods in multi-image unsupervised denoising\citep{batson2019noise2self,Krull2019}, we investigate a self-supervised neural networks Self2Self with the following basic form: 
\begin{equation}\label{eq:loss_single-self}
\underset{\vtheta}{\min}\ \ell(\oF_{\vtheta}(\hat{\vy}),\hat{\vy}).
\end{equation} 
where $\hat{\vy}$ denotes counterpart images that are generated from the corrupted image $\vy$, and $\oF_\vtheta(\cdot)$ represents the neural networks.

Compared to the multi-image supervised/unsupervised ones, an effective single-image self-supervised recovery approach is much more challenging to develop. The learned model from single-image supervision is probably over-fit and biased as the training data amount is very limited. While a single model may have bias, the ensemble of multiple NNs with different model configurations may cancel the bias out. Nevertheless, there are still three key problems needed to solve when using NN ensemble for our task:\\
{\it - [P1] How to train and maintain multiple models economically? Separately training a number of NNs is impractical.}\\
{\it - [P2] How to improve the effectiveness of each learned model? A model supervised by a single corrupted image may just memorize the target or learn identity mapping.}\\
{\it - [P3] How to ensure the diversity of the learned models so that their ensemble can bring noticeable improvement?}

In this paper, we solve all of the above problems by dropout~\citep{srivastava2014dropout}, which refers to dropping out nodes in an NN randomly.
We construct an NN with dropout layers and use dropout during training as well as test of the NN:
\begin{itemize}[leftmargin=15pt]
	\setlength{\itemsep}{0pt}
	\setlength{\parsep}{0pt}
	\setlength{\parskip}{0pt}
	\setlength{\topsep}{0pt}
	\item During training, the dropout layers enable a single NN approximating a large number of different NN architectures by randomly dropping out nodes. Thus, it offers a very computationally cheap method to train multiple models. \ {\it[Ans. to P1]}
	\item Dropout naturally prevents the NN overfitting on limited amount data, it reduces the possibility of purely memorizing the single target and learning trivial mappings. Dropout is also applied to the input layer, which equals to randomly dropping pixels on the input image for preventing learning identity mapping. Together with partial convolutions~\citep{liu2018image} that ignore the effect of masked pixels during restoration, our dropout NN model can learn single-image recovery effectively. \ {\it [Ans. to P2]}
	\item In test, we use dropout again on the trained NN to generate multiple models for image recovery. It is shown that the model trained by dropout has high model uncertainty and the dropout models generated from it during test have sufficient diversity. \	{\it [Ans. to P3]}
\end{itemize}

\subsection{Training} 
In training, a series of counterpart images $\hat{\vy}_1,\cdots,\hat{\vy}_M$ are generated from the given corrupted image $\vy$ as follows:
\begin{equation}
\hat{\vy}_m=\mB_m\vy, \ \ \text{for all}\ m,
\end{equation}
where $\mB_m$ is a random binary diagonal matrix with its diagonal entry $\mB_m(i,i)$ drawn from a Bernoulli distribution with probability $p$.
Then, we train our NN model $\oF_\vtheta(\cdot)$ to map each $\hat{\vy}_m$ to  $\vy$, with dropout and the loss functions:
\begin{equation}
\label{common-inpainting-loss}
\ell(\vtheta)=\sum_{m=1}^M \left \|(\mI-\mB_m) (\oF_{\vtheta}(\hat{\vy}_m)-\vy)\right \|^2_2.
\end{equation}
Note $\mI-\mB$ implies that the loss is defined on the dropped pixels.  For improvement, data augmentation is used by flipping the given image horizontally, vertically and diagonally. For computation efficiency,  when noisy images from the same dataset are strongly relevant to each other, we first train base NN parameters for this dataset, then separately fine-tune each image via updating the parameters from the base one.

\subsection{Test}
In the test stage, multiple new NNs ($\oF_{\vtheta_1},\cdots,\oF_{\vtheta_N}$) are formed by randomly running dropout on the configured layers of the trained NN $\oF_{\vtheta^*}$. Then, multiple recovered images $\hat{\vx}_1,\cdots,\hat{\vx}_N$ are generated by feeding a radomly re-corrupted version of $\vy$ to each of the newly-formed NNs. The recovered images are then averaged to obtain the final result $\vx^*$:
\begin{equation}
	\vx^*=\sum_{n=1}^N \hat{\vx}_i=\sum_{n=1}^N \oF_{\vtheta_n}(\mB_{M+n}\vy).
\end{equation}

\subsection{NN architecture}

The NN architecture is borrowed from UNet\citep{Ronneberger2015} with some modifications. See Supplementary Information for an illustration. There are mainly two differences between UNet network and ours. First, we introduce dropout to the Conv layers.  In a layer with dropout, there is a probability for each weight entry that will be set as zero, and those remaining entries would be scaled for energy maintaining. The dropout layers are used in both training and testing, which is for regularizing the NN, preventing learning trivial mappings and learning multiple diverse NNs simultaneously in an economic way. Second, we use partial convolutions instead of the standard ones in the encoder part. This can eliminate the effect of the dropped pixels during the processing, with both effectiveness and efficiency improvement observed in practice.

\bmhead{Data availability}
The HMI LOS magnetograms and HST galaxy images involved in this paper are downloaded in the Joint Science Operations Center (http://jsoc.stanford.edu/) and Mikulski Archive for Space Telescopes ( https://archive.stsci.edu/). Interested parties can also download the corresponding data at the project page (https://github.com/zimugh/Denoising-TDR/).

\bmhead{Code availability}
The codes of the TDR-method are publicly available. We provide a Python notebook to show the denoising examples at https://github.com/zimugh/Denoising-TDR/.

\bmhead{Correspondence and requests for materials}
should be addressed to 
T.L., Y.Q., Y.S. or Y.G. 

\bmhead{Acknowledgments}
The magnetograms are observed by the Helioseismic Magnetic Imager onboard the Solar Dynamics Observatory which is a mission of Living With a Star Program from NASA. The galaxy images are based on observations made by the NASA/ESA Hubble Space Telescope obtained from the Space Telescope Science Institute, which is operated by the Association of Universities for Research in Astronomy, Inc., under NASA contract NAS 5–26555. These galaxy observations are associated with program(s) 5222. The neural network Self2Self is constructed by Tensorflow. This work is supported by the National Key R\&D Program of China 2022YFF0503004(T.L., Y.G., Y.W.), 2021YFA1600502(Y.S.), 2020YFC2201201(Y.G.), 2022YFF0503001(Y.S.), by the National Natural Science Foundation of China (NSFC) Grant No. 12203023(T.L.), 62372186(Y.Q), 12333009(Y.G.), 12173092(Y.S.), by the Natural Science Foundation of Guangdong Province Grant No. 2022A1515011755(Y.Q.), 2023A1515012841(Y.Q.), by Fundamental Research Funds for the Central Universities Grant No. 2023ZYGXZR022(Y.Q.), and by the Strategic Priority Research Program of the Chinese Academy of Sciences Grant No. XDB0560000(Y.S., H.J.). 

\bmhead{Author contributions}
T.L., Y.Q., Y.S. and Y.G. led the study, developed the method and wrote the manuscript. S.L, H.J. and Q.H. contributed to providing the scientific data and the analysis of the denoising effects on the magnetograms. Y.L.G. performed the analysis of the application to Hubble images. Y.L., Y.W., W.S. and M.D. added the test on synthetic images, improved the definition of the accuracy and polished the manuscript. All authors contributed to the discussion and improving the manuscript.

\bmhead{Competing interests}
The authors declare no competing interests.

\bmhead{Additional information}
Supplementary Information is available in a separate file.

\section[Supplementary Information]{Supplementary Information}
\subsection{Supplementary tables}\label{supp}

\begin{sidewaystable}
    \renewcommand{\tablename}{Supplementary Table}
	\centering
	\footnotesize
	\caption{The quantitative evaluation parameters. The computing formulas are defined in Supplementary Information \ref{secB}. Nine paired images are obtained to calculate the mean values and standard deviations. Four decimal places are uniformly retained, note that the numerical accuracy in the table only represents statistical significance and does not represent image accuracy.} 
    \begin{tabular}{cccc}
    \toprule
	Quantitative Evaluation Parameters & MV $\pm$ SD (Denoised) &  MV $\pm$ SD (Denoised + R10) &  MV $\pm$ SD (Denoised + R5)\\
    \midrule
	Absolute Error (AE)  & -0.2330 $\pm$ 0.0200 & -0.1302 $\pm$ 0.0138 & -0.0361 $\pm$ 0.0039 \\
	\hline
	Mean Square Error (MSE) &0.0246 $\pm$ 0.0050 & 0.0074 $\pm$ 0.0019 & 0.0015 $\pm$ 0.0004\\
	\hline
	Pearson Correlation Coefficient (PCC) & 0.9891 $\pm$ 0.0022 & 0.9963 $\pm$ 0.0009 & 0.9992 $\pm$ 0.0002\\
	\hline	
	Linear Coefficient (LC) & 0.9243 $\pm$ 0.0099 & 0.9910 $\pm$ 0.0023 & 0.9982 $\pm$ 0.0005\\
	\hline	
	Average flux of residual image (AFR) & -0.0199 G $\pm$ 0.0772 G & 0.0131 G $\pm$ 0.0072 G & 0.0039 G $\pm$ 0.0023 G \\
	\hline	
	Average absolute flux of residual image (AAFR) & 5.6644 G $\pm$ 0.0974 G & 3.4864 G $\pm$ 0.0111 G & 1.3132 G $\pm$ 0.0065 G \\
    \hline	
	Noise Level: Initial HMI LOS magnetograms & 8.0802 G $\pm$ 0.0591 G & 8.0802 G $\pm$ 0.0591 G & 8.0802 G $\pm$ 0.0591 G \\
    & $\Downarrow$ & $\Downarrow$ & $\Downarrow$ \\
	Noise Level: Denoised HMI LOS magnetograms & 2.0738 G $\pm$ 0.0170 G & 2.0486 G $\pm$ 0.0156 G & 2.1185 G $\pm$ 0.0154 G \\
	\bottomrule

	\label{tab: cha}
	\end{tabular}
\end{sidewaystable}

\begin{sidewaystable}
    \renewcommand{\tablename}{Supplementary Table}
	\centering
	\footnotesize
	\caption{NN architecture in proposed method. LReLU is for Leaky ReLU. Parameters: dropout ratio for Dropout, receptive field size for Max Pooling, scaling factor for Upsampling, and two block sequence numbers (outputs of which blocks to concatenate) for Concatenation.}\vspace{5pt}
	\begin{tabular}{cccc|cccc}
		\toprule
		NO.    & Function  & \# Channels & Parameters     & NO.    & Function  & \# Channels  & Parameters \\
		\midrule
		\multicolumn{4}{c|}{ENCODER}  & 17    & Dropout + Conv. + LReLU   & 96    & 30\% \\
		\cmidrule{1-4}    1     & -     & C     & -     & 18    & Upsampling & 96    & x 2 \\
		2     & PConv. + LReLU   & 48    & 30\% & 19    & Concatenation & 144   & 8 + 18 \\
		3     & PConv. + LReLU   & 48    & 30\% & 20    & Dropout + Conv. + LReLU   & 96    & 30\% \\
		4     & Max Pooling & 48    & 2 x 2 & 21    & Dropout + Conv. + LReLU   & 96    & 30\% \\
		5     & PConv. + LReLU   & 48    & 30\%  & 22    & Upsampling & 96    & x 2 \\
		6     & Max Pooling & 48    & 2 x 2 & 23    & Concatenation & 144   & 6 + 22 \\
		7     & PConv. + LReLU   & 48    & 30\% & 24    & Dropout + Conv. + LReLU   & 96    & 22.5\% \\
		8     & Max Pooling & 48    & 2 x 2 & 25    & Dropout + Conv. + LReLU   & 96    & 30\% \\
		9     & PConv. + LReLU   & 48    & 30\%  & 26    & Upsampling & 96    & x 2 \\
		10    & Max Pooling & 48    & 2 x 2 & 27    & Concatenation & 144   & 4 + 26 \\
		11    & PConv. + LReLU   & 48    & 30\% & 28    & Dropout + Conv. + LReLU   & 96    & 30\% \\
		12    & Max Pooling & 48    & 2 x 2 & 29    & Dropout + Conv. + LReLU   & 96    & 30\% \\
		13    & PConv. + LReLU   & 48    & 30\%  & 30    & Upsampling & 96    & x 2 \\
		\cmidrule{1-4}    \multicolumn{4}{c|}{DECODER}  & 31    & Concatenation & 96 + C  & 1 + 30 \\
		\cmidrule{1-4}    14    & Upsampling & 48    &  x 2 & 32    & Dropout + Conv. + LReLU   & 64    & 30\% \\
		15    & Concatenation & 96    & 10 + 14 & 33    & Dropout + Conv. + LReLU   & 32    & 30\% \\
		16    & Dropout + Conv. + LReLU   & 96    & 30\% & 34    & Dropout + Conv. + ReLU   & C     & 30\% \\
		\bottomrule
	\end{tabular}
	\label{tab:architecture}
\end{sidewaystable}

In Supplementary Table \ref{tab: cha}, the mean value of AE is $-0.2330$ for `Denoised' magnetograms, $-0.1302$ for `(Denoised+R10)' magnetograms and $-0.0361$ for `(Denoised+R5)' magnetograms, respectively, which is mostly caused by the reduction of the noises decreasing the absolute flux shown by the AAFR of $\mathrm{5.6644 \pm 0.0974}$ G, $\mathrm{3.4864 \pm 0.0111}$ G, $\mathrm{1.3132 \pm 0.0065}$ G for the `Denoised', `(Denoised+R10)', `(Denoised+R5)' magnetograms. Although the noise reduction is remarkable, the deviation of denoised magnetograms from the initial magnetograms is small, with the mean values of MSE, PCC, LC equal to $0.0246, \ 0.9891, \ 0.9243$ for `Denoised' magnetograms, $0.0074, \ 0.9963, \ 0.9910$ for `(Denoised+R10)' magnetograms and $0.0015, \ 0.9992, \ 0.9982$ for `(Denoised+R5)' magnetograms, respectively, which indicates that the two images before and after denoising are highly consistent with each other. The mean value of AFR is close to zero, as $-0.0199$ G for `Denoised' magnetograms, $0.0131$ G for `(Denoised+R10)' magnetograms and $0.0039$ G for `(Denoised+R5)' magnetograms, which indicates that the denoised magnetograms keep the same total flux as the initial magnetograms and zero-mean noise is displayed in the residual images. The standard deviations of AE, MSE, PCC, LC, AFR and AAFR are all small enough, this suggests that the TDR-method is robust and reliable.

Basically, our NN is an encoder-decoder network, shown in Supplementary Table \ref{tab:architecture}.
Given an input image of the size $H\times W\times C$, the encoder part first maps the image to an $H\times W\times 48$ feature cube with a partial convolutional (PConv) layer,  which is then processed by the following six encoder's blocks (EBs). Each of the first five EBs sequentially connects a PConv layer, a leaky rectified linear unit (LReLU), and a max pooling layer with $2\times2$ receptive fields and with a stride of $2$. The last block only contains a PConv layer and an LReLU. The number of channel is fixed to $48$ across all the encoder's blocks. As a result, the output of the encoder is a feature cube of size $H/32\times W/32 \times 48$.
The decoder part contains five decoder's blocks (DBs). Each of the first five DBs sequentially connects an upsampling layer with a scaling factor of $2$, a concatenation operation, and two standard convolutional (Conv) layers with dropout and followed by LReLUs. The concatenation operation in a DB stacks together the feature cube from the upsampling layer and the one output by the LReLU in the corresponding EB, which utilize more information. All the Conv layers in the first four DBs have $96$ output channels. The last DB contains three Conv layers with LReLU/ReLU for mapping the feature cube back to an image of size $H\times W\times C$, and the numbers of output channels of its three Conv layers are $64,32,C$ respectively.

\subsection{Data and evalution parameters}\label{secB}

The magnetograms are measured by the Helioseismic Magnetic Imager (HMI) onboard the Solar Dynamics Observatory (SDO) lunched on February 11, 2010. HMI measures the magnetic field on the solar photosphere and provides line-of-sight (LOS) magnetograms of the Sun with cadence of 45 seconds and spatial resolution of $0.''5$ per pixel. Ten HMI LOS magnetograms (from 2013-01-01 00:01:30 UT to 2013-01-10 
00:01:30 UT) with cadence of one day and size of $4096 \times 4096$ pixels are collected. In order to obtain images of appropriate size and the best observation quality, the central subsets with the size of $1024 \times 1024$ pixels from the whole magnetograms are cut out and used in this work. 

We obtain images about the central region of the galaxy NGC 1365 by two wide filters of F555W and F814W from the Hubble Space Telescope (HST). The two filters of F555W and F814W correspond to the central wavelengths of about 534.6 and 833.3 nm. First, three images of F555W and three images of F814W with the exposure time of 10 s are obtained. And then the central subsets with the size of $300 \times 300$ pixels are denoised by the TDR-method. To verify the noise reduction effect, the denoised images are compared with the corresponding images with the exposure time of 100 s. 

\begin{equation}
\label{ae}
AE_i=  \sum\left(\mid\chi_{j}^{d}\mid -\mid\chi_{j}^{o}\mid\right ) / \sum\mid\chi_{j}^{o}\mid,
\end{equation}

\begin{equation}
\label{mse}
MSE_i=  \sum\left(\chi_{j}^{d} -\chi_{j}^{o}  \right )^2 /\sum \left ( \chi_{j}^{o} \right )^2,
\end{equation}

\begin{equation}
\label{pcc}
PCC_i= \frac{\sum\left ( \chi_{j}^{o}-m_\chi^{o} \right )\left ( \chi_{j}^{d}-m_\chi^{d} \right )}{\sqrt{\sum\left ( \chi_{j}^{o}-m_\chi^{o} \right )^2 \sum\left ( \chi_{j}^{d}-m_\chi^{d} \right )^2}},
\end{equation}

\begin{equation}
\label{afr}
AFR_i=\sum\left(\chi_{j}^{o} -\chi_{j}^{d}\right ) / \sum j,
\end{equation}

\begin{equation}
\label{aafr}
AAFR_i=\sum\mid\chi_{j}^{o} -\chi_{j}^{d}\mid / \sum j,
\end{equation}

The computing formulas of AE, MSE, PCC, AFR and AAFR in Supplementary Table \ref{tab: cha} are listed above, where $i$ is the serial number of the evaluation samples and $j$ represents the pixel number on an image. $\chi_{j}^{d}$ and $\chi_{j}^{o}$ denote the pixel values of the denoised and original images, repectively. $m_\chi^{d}$ and $m_\chi^{o}$ are the mean values of the $\chi_{j}^{d}$ and $\chi_{j}^{o}$ over an image. We calculate the standard deviations of the evaluation scores in Supplementary Table \ref{tab: cha} by equation of $SD=\sqrt{\sum^{n}_{i=1}\left ( s_i-m_s \right )^2/n}$, where $m_s$ denotes the mean value of the score $s_i$.

\subsection{Test on synthetic images}\label{secC}

\begin{figure*}
    \renewcommand{\figurename}{Supplementary Figure}
	\centering \includegraphics[width=\textwidth]{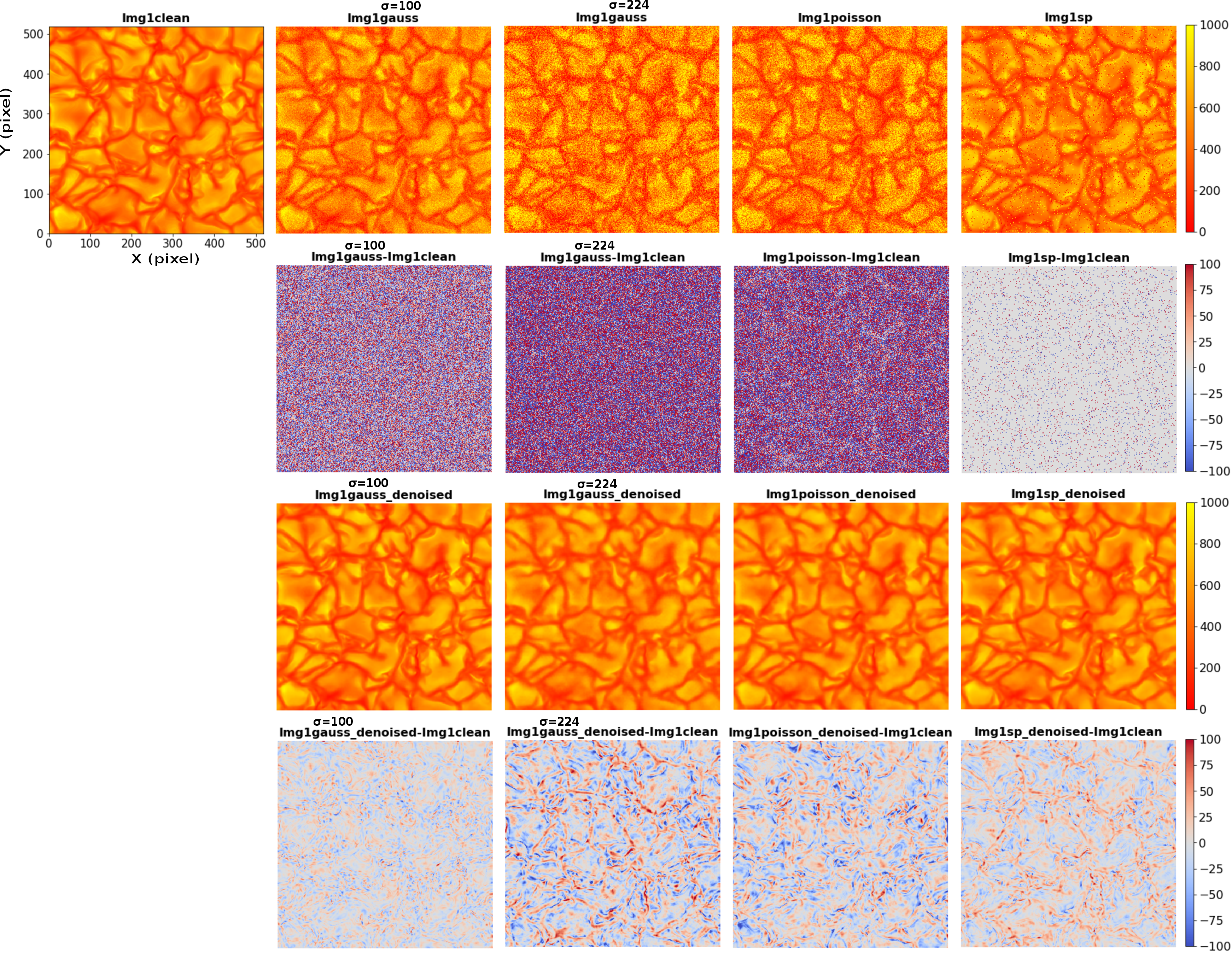}
	\caption{Noise reduction of the synthetic images. The clean image is contaminated by Gaussian noises with the standard deviations $\sigma=100$ and $\sigma=224$  as well as Poisson noise and salt-and-pepper noise. The first row shows the clean target image and the corresponding images with different noises. The residual maps of the noisy images and the clean target image are displayed in the second row. The third and last rows show the corresponding denoised images and residual maps of the dnoised images and the clean target image.} 	
	\label{fig:f7}  
\end{figure*}

\begin{figure*}
    \renewcommand{\figurename}{Supplementary Figure}
	\centering \includegraphics[width=\textwidth]{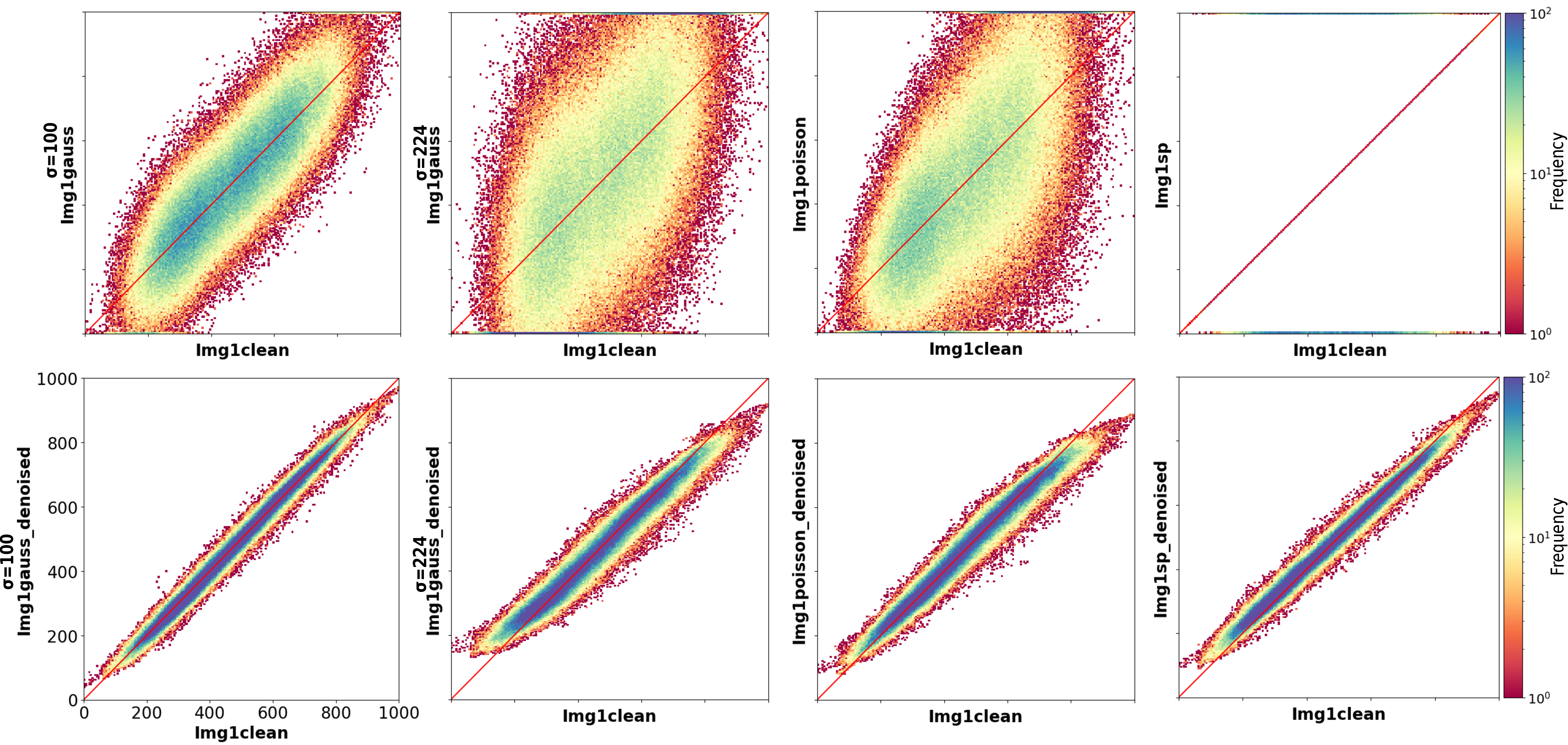}
	\caption{The density-scatter plots of images in Supplementary Figure \ref{fig:f7}. The first row shows the density-scatter plots of the noisy images and clean target image, and the bottom row shows those of the denoised images. Note that the salt-and-pepper noise is constructed by randomly replacing 5 \% of the initial pixels with the value of 0 or 1000 (maximum or minimum value). As a result, 95 \% pixels are the initial pixels, exactly follow the 1-1 relation and are blocked by the red line, the other 5 \% pixels are the added salt-and-pepper noises and located at the top and bottom in the fourth panel.}
	\label{fig:f8}  
\end{figure*}

\begin{table*}
    \renewcommand{\tablename}{Supplementary Table}
	\centering
	\footnotesize
	\caption{The evaluation parameters of Gauss and Poisson noises. Pixels in the clean target images are considered as true, and the MSE, AFR, AAFR and PCC scores of the noisy and denoised images are calculated. Five groups of images are obtained to calculate the mean values and standard deviations. Four decimal places are uniformly retained, note that the numerical accuracy here only represents statistical significance and does not represent image accuracy.} 
    \begin{tabular}{ccc}
    \toprule
	 Gauss $\sigma=100$ & MV $\pm$ SD (Img) &  MV $\pm$ SD (Img\_denoised) \\
    \midrule
	MSE &0.0457 $\pm$ 0.0049 & 0.0018 $\pm$ 0.0003 \\
	\hline	
	AFR & 0.2755 $\pm$ 0.2081 & 0.1571 $\pm$ 2.0028 \\
	\hline	
	AAFR & 79.5628 $\pm$ 0.0527 & 15.0406 $\pm$ 0.3180 \\
 	\hline
	PCC & 0.8476 $\pm$ 0.0109 & 0.9925 $\pm$ 0.0010 \\
	\bottomrule
     &  &  \\

	 Gauss $\sigma=224$ & MV $\pm$ SD (Img) &  MV $\pm$ SD (Img\_denoised) \\
    \midrule
	MSE &0.2024 $\pm$ 0.0213 & 0.0055 $\pm$ 0.0007 \\
	\hline	
	AFR & 3.8990 $\pm$ 1.8331 & 2.5888 $\pm$ 3.7091 \\
	\hline	
	AAFR & 170.0964 $\pm$ 0.4792 & 26.6798 $\pm$ 0.3248 \\
 	\hline
	PCC & 0.5821 $\pm$ 0.0178 & 0.9782 $\pm$ 0.0023 \\
    \bottomrule
     &  &  \\

Poisson  & MV $\pm$ SD (Img) &  MV $\pm$ SD (Img\_denoised) \\
    \midrule
	MSE &0.1497 $\pm$ 0.0097 & 0.0041 $\pm$ 0.0003 \\
	\hline	
	AFR & -1.3227 $\pm$ 1.1255 & -1.7524 $\pm$ 0.8090 \\
	\hline	
	AAFR & 143.4949 $\pm$ 3.3866 & 22.5405 $\pm$ 0.5978 \\
 	\hline
	PCC & 0.6527 $\pm$ 0.0075 & 0.9834 $\pm$ 0.0014 \\
    \bottomrule
	\end{tabular}
    \label{tab: cha1}
\end{table*}	

\begin{table*}
    \renewcommand{\tablename}{Supplementary Table}
	\centering
	\footnotesize
	\caption{Percentage of pixels in five ranges. P represents the absolute value of the change (compared with the clean target images) in pixel value for the noisy and denoised images. The percentages of pixels with change values in five ranges are listed. Five groups of images are obtained to calculate the mean values and standard deviations.} 
    \begin{tabular}{ccc}
    \toprule
	Gauss $\sigma=100$ & MV $\pm$ SD (Img) &  MV $\pm$ SD (Img\_denoised) \\
    \midrule
	100 < P & (31.70 $\pm$ 0.09) \% & (0.04 $\pm$ 0.02) \% \\
	\hline	
 
	50 $<$ P $\leq$ 100   & (30.03 $\pm$ 0.07) \% & (1.93 $\pm$ 0.30) \% \\
	\hline	
 
	20 $<$ P $\leq$ 50 & (22.43 $\pm$ 0.03) \% & (25.17 $\pm$ 0.50) \% \\
 	\hline

	10 $<$ P $\leq$ 20 & (7.90 $\pm$ 0.02) \% & (29.95 $\pm$ 0.32) \% \\
 	\hline
  
	P $\leq$ 10  & (7.94 $\pm$ 0.05) \% & (42.90 $\pm$ 0.75) \% \\
	\bottomrule
    &  &  \\

	Gauss $\sigma=224$ & MV $\pm$ SD (Img) &  MV $\pm$ SD (Img\_denoised) \\
    \midrule
	100 < P & (65.32 $\pm$ 0.14) \% & (0.86 $\pm$ 0.05) \% \\
	\hline	
 
	50 $<$ P $\leq$ 100  & (16.91 $\pm$ 0.07) \% & (12.91 $\pm$ 0.27) \% \\
	\hline	
 
	20 $<$ P $\leq$ 50  & (10.61 $\pm$ 0.06) \% & (39.16 $\pm$ 0.48) \% \\
 	\hline

	10 $<$ P $\leq$ 20  & (3.57 $\pm$ 0.05) \% & (22.15 $\pm$ 0.29) \% \\
 	\hline
  
	P $\leq$ 10  & (3.59 $\pm$ 0.01) \% & (24.92 $\pm$ 0.46) \% \\
	\bottomrule
    &  &  \\

	Poisson  & MV $\pm$ SD (Img) &  MV $\pm$ SD (Img\_denoised) \\
    \midrule
	100 < P & (57.23 $\pm$ 1.07) \% & (0.64 $\pm$ 0.10) \% \\
	\hline	
 
	50 $<$ P $\leq$ 100   & (20.41 $\pm$ 0.46 \% & (8.18 $\pm$ 0.65) \% \\
	\hline	
 
	20 $<$ P $\leq$ 50  & (13.26 $\pm$ 0.34) \% & (35.87 $\pm$ 0.44) \% \\
 	\hline

	10 $<$ P $\leq$ 20  & (4.52 $\pm$ 0.13) \% & (25.25 $\pm$ 0.35) \% \\
 	\hline
  
	P $\leq$ 10  & (4.59 $\pm$ 0.15) \% & (30.07 $\pm$ 0.71) \% \\
	\bottomrule
    
	\end{tabular}
    \label{tab: cha2}
\end{table*}	

To show the denoising effect of the neural network Self2Self more clearly, we adopt five synthetic images from a simulation carried out by RAMENS code as the clean target images, which show the simulated granulation elements on the solar photosphere. First, the pixel values of the synthetic images are renormalized to the range of [0, 1000]. Second, Gaussian noises with the standard deviation $\sigma=100$ and $\sigma=224$ as well as Poisson noise and salt-and-pepper noise are added to the clean target images. As a result, five clean-noisy image groups are obtained. The first example of the five groups is shown in the first row of Supplementary Figure \ref{fig:f7}. The added noises are displayed by the residual maps in the second row. The distribution of Gaussian noise is independent with the structures in the clean target image, while Poisson noise is proportional to the intensities of the granulation elements. The salt-and-pepper noise is constructed by randomly replacing 5 \% of the initial pixels with the value of 0 or 1000.

We choose the first image with Gaussian noise ($\sigma=100$) to train the neural network Self2Self and obtain a denoising model which is used to denoise the four different noisy images in the first row of Supplementary Figure \ref{fig:f7}. One example group of the denoised images and corresponding residual maps are displayed in the third and bottom rows of Supplementary Figure \ref{fig:f7}. The residual maps in the second row show the added noises, and those in the bottom row show the denoising results. We find that most of the Gaussian noise has been removed. Pixels with large residual values are concentrated in the gaps between granulation elements, due to the lower signal-to-noise ratios of these pixels. Compared with Gaussian noise, the noise reduction effect of Poisson noise is better, due to the lower noise intensity in the gaps between granulation elements. From the example of Gaussian noises ($\sigma=100$ and $\sigma=224$) and Poisson noise, we find that the noise reduction effect of the denoising model is positively correlated with the signal-to-noise ratios of pixels on an image. This is consistent with our intuitive understanding about the noise and signal. The more severely the signal is polluted by noise, the more difficult the signal is recovered. The example of salt-and-pepper noise might simulate image damage caused by defects of observation instruments. The four examples of noise reduction indicate that our method can deal with various forms of noises, which may exist simultaneously in astronomical images.

The density-scatter plots in Supplementary Figure \ref{fig:f8} show the level of compatibility between the noisy, denoised images and the clean target image. A common observation is that pixels with values in proximity to the maximum and minimum tend to show more significant deviations from the true values. This can be attributed to the smoothing effect of noise reduction by Self2Self, and be suppressed by the restoring operation for observation images. We further quantify the differences between the noisy, denoised images and the clean target image. The rusults are listed in Tables \ref{tab: cha1} and \ref{tab: cha2}, where the example of salt-and-pepper noise is not included due to that its special noise distribution shown in the fourth panel of Supplementary Figure \ref{fig:f8} makes the evaluation parameters ineffective to measure its denoising  effect. By comparing the scores of the other three examples of Gaussian noises ($\sigma=100$ and $\sigma=224$) and Poisson noise, we find that a significant portion of the noise has been eliminated, resulting in a high degree of consistency between the denoised images and the clean target images. Although the denoised images cannot be entirely restored to their corresponding clean target images, the distribution of pixel changes in Supplementary Table \ref{tab: cha2}, indicates that denoising by Self2Self brings contaminated pixels much closer to their true values. After denoising, the percentage of pixels with a change amount exceeding 100 (the minimum noise level among the three examples) is below 1 \% for the three examples in Supplementary Table \ref{tab: cha2}. Before denoising they are about 31.70 \%, 65.32 \% and 57.23 \%, respectively, which means that a statistically significant portion of weak signals is restored. This proves the reliability of the denoising method.



\end{document}